\documentstyle[12pt]{article}
\begin{document} 

\begin{center}
{\bf Quantum Hall effect anomaly and collective modes in the
  magnetic-field-induced spin-density-wave phases of 
  quasi-one-dimensional conductors}
\end{center}
\vspace{1cm}


N. Dupuis\footnote{On leave from Laboratoire de Physique des Solides,
  Universit\'e Paris-Sud, 91405 Orsay, France} and Victor M. Yakovenko 
\vspace{1cm}

Department of Physics and Center for Superconductivity Research, 

University of Maryland, 
College Park, MD 20742-4111, USA  
\vspace{2cm}


\vspace{2cm}
We study the collective modes in the magnetic-field-induced
spin-density-wave (FISDW) phases experimentally observed in organic
conductors of the Bechgaard salts family. In phases that exhibit a sign
reversal of the quantum Hall effect (Ribault anomaly), the
coexistence of two spin-density waves gives rise to additional long
wavelength collective
modes besides the Goldstone modes due to spontaneous translation and
rotation symmetry breaking. These modes strongly affect the charge and spin
response functions. We discuss some experimental consequences for the
Bechgaard salts.

\vspace{1cm}

\noindent
{\it Introduction.} - 
The organic conductors of the Bechgaard salts family (TMTSF)$_2$X (where
TMTSF stands for tetramethyltetraselenafulvalene) have remarkable properties
in a magnetic field. In three members of this family (X=ClO$_4$, PF$_6$,
ReO$_4$), a moderate magnetic field of a few Tesla destroys the
metallic phase and induces a series of SDW phases
separated by first-order phase transitions \cite{rev,Heritier84}.

According to the so-called quantized nesting model (QNM)
\cite{Heritier84}, the formation of the FISDWs results from the
strong anisotropy of these organic materials, which can be viewed as weakly
coupled chain systems (the typical ratio of the electron
transfer integrals in the three crystal directions is
$t_a:t_b:t_c=3000:300:10$ K). The SDW opens a gap, but leaves
closed pockets of electrons and/or holes in the vicinity of the Fermi
surface. In presence of a magnetic field $H$, these pockets are
quantized into Landau levels (more precisely Landau subbands). In each FISDW
phase, the SDW wave vector is quantized, ${\bf Q}_N=(2k_F+NG,Q_y)$ with $N$
integer, so that an integer number of Landau subbands are filled. (Here $k_F$
is the Fermi momentum along the chains, $-e$ the electron charge, $b$ the
interchain spacing, and $G=eHb/\hbar $.) As a result, the Fermi level lies
in a gap between two Landau subbands, the SDW phase is stable, and the Hall
conductivity is quantized: $\sigma _{xy}=-2Ne^2/h$ per one layer of the
TMTSF molecules \cite{Poilblanc87,Yakovenko91}. As the magnetic field
increases, the value of the integer $N$ changes, which leads to a cascade of
FISDW transitions. 

A striking feature of the QHE in Bechgaard salts is the coexistence of both
positive and negative Hall plateaus. While most plateaus are of the same
sign, referred to as positive by convention, a negative Hall effect is also
observed at certain pressures (the so-called Ribault anomaly) 
\cite{Ribault85}.  

We have recently explained the Ribault anomaly within the framework of the
QNM by taking umklapp processes into account \cite{ND}. Because of
umklapp scattering, two linearly polarized SDWs, with wave vectors ${\bf
Q}_N=(2k_F+NG,\pi /b)$ and ${\bf Q}_{-N}=(2k_F-NG,\pi /b)$, coexist in the
Ribault phase \cite{coex}. The quantum Hall conductivity is quantized:
$\sigma _{xy}=-2Ne^2/h$. The integer $N$ is negative (hence a negative Hall
plateau) and corresponds to the SDW with the largest
amplitude. Experimentally, $N=-2$ and $N=-4$ are the most commonly observed
`negative' phases. Our explanation of the Ribault anomaly differs from
the one suggested by Zanchi and Montambaux \cite{Zanchi96} by invoking the pressure
dependence of umklapp scattering rather than the electron band
structure. 

In this Letter, we study the long wavelength collective
modes in the FISDW phases that exhibit the Ribault 
anomaly. The coexistence of two SDWs in these phases \cite{coex} gives rise
to additional collective modes besides the Goldstone modes resulting from
spontaneous rotation and translation symmetry breaking. We point out some 
analogies with phase modes in two-band or bilayer superconductors
\cite{Legett66,Wu95} and plasmon modes in semiconductor double-well
structure \cite{Dassarma81}. While these modes are generally difficult to
observe, collective modes have strong experimental consequences in SDW
systems \cite{Gruner}. We discuss how the charge and spin response functions
are  affected in the Ribault phase.  

It should be pointed out that the long wavelength modes are not the
only modes of interest in the FISDW phases. There also exist
magneto-rotons at finite wave vectors ($q_x=G,2G,...$)
\cite{Poilblanc87a}. Within the
Zanchi-Montambaux scheme, Lederer has recently shown that these modes
exhibit a different behavior in the Ribault phase
\cite{Lederer98}. The effect of umklapp scattering on the
magneto-rotons is not considered in this Letter. 

\vspace{.3cm}
\noindent
{\it Mean-field theory.} - 
In the vicinity of the Fermi energy, the electron dispersion law in the
Bechgaard salts is approximated as  
\begin{equation}
  E(k_x,k_y) = v_F(|k_x| -k_F) + t_\perp(k_yb),
\label{disp}
\end{equation}
where $k_x$ and $k_y$ are the electron momenta along and across the
one-dimensional chains of TMTSF, and $\hbar=1$. In Eq.\ (\ref{disp}),
the longitudinal electron dispersion is linearized in $k_x$ in the
vicinity of the two one-dimensional Fermi points $\pm k_F$, and
$v_F=2at_a\sin(k_Fa)$ is the corresponding Fermi velocity ($a$ is the
lattice spacing along the chains). The periodic
function $t_\perp (u)=t_\perp (u+2\pi)$ describes the interchain hopping in
a tight-binding approximation:
\begin{equation}
t_\perp (k_yb) = -2t_b\cos(k_yb)-2t_{2b}\cos(2k_yb)
-2t_{3b}\cos(3k_yb)-2t_{4b}\cos(4k_yb).
\end{equation}
We neglect the electron
dispersion in the third direction along the $z$ axis. $t_{4b}$
plays a crucial role in the 
Ribault phase since (together with the umklapp scattering strength) it
determines the ratio of the amplitudes of the two SDWs. For
$t_{4b}=0$, both SDWs have the same amplitude (independently of the
value of $t_{3b}$). A finite $t_{4b}$ lifts
this degeneracy. This yields a negative QHE whenever ${\rm
sgn}(t_{4b})=\rm{sgn}(t_{2b})$ \cite{ND}. On the other hand, $t_{3b}$ does
not affect directly the Ribault phase 
\cite{note1}. Its only effect is to change the critical value of
umklapp scattering above which the Ribault phase becomes stable. 
As shown by Zanchi and Montambaux \cite{Zanchi96}, a sufficiently strong
$t_{3b}$ ($t_{3b} \stackrel{\textstyle >}{\sim } 0.2 t_{2b}$) 
can stabilize a negative phase even in the absence of umklapp
scattering. We cannot exclude that both umklapp scattering and
dispersion law play a role in stabilizing the Ribault phase at low
pressure. However, whatever the mechanism at work, umklapp scattering (not
considered in the Zanchi-Montambaux theory) will always lead to the
coexistence of two SDWs in the Ribault phase. 
 
The presence of a magnetic field along the $z$ direction is taken into
account via the Peierls substitution ${\bf k}=(k_x,k_y)\to -i
\mbox{\boldmath$\nabla$}
-e{\bf A}$ where $\bf A$ is the vector potential. The magnetic field also
introduces the Zeeman coupling $\sigma \mu _BH$ where $\sigma =+(-)$ for up
(down) spins. ($\mu _B$ is the Bohr magneton. We take the electron 
gyromagnetic factor $g$ equal to two.)  
We consider forward and umklapp scatterings between electrons, with
amplitudes $g_2$ and $g_3$ respectively. We do not consider backward
scattering, since it does not play an important role in the QNM \cite{ND}. 

In the FISDW phase, the electron spin density has a nonzero expectation
value:
\begin{eqnarray}
\langle S_x({\bf r})\rangle &=&
  \sum_{\beta =\pm} m_{\beta N} \cos (\phi _{\beta N})  
  \cos( {\bf r}\cdot{\bf Q}_{\beta N} + \theta_{\beta N}),
\nonumber \\ 
\langle S_y({\bf r})\rangle &=& 
   \sum_{\beta =\pm} m_{\beta N} \sin (\phi _{\beta N})  
  \cos( {\bf r}\cdot{\bf Q}_{\beta N} + \theta_{\beta N}),
\label{sdsin}
\end{eqnarray}
where ${\bf r}=(x,y)$ is the spatial coordinate.
Because of the Zeeman coupling with the magnetic field, the SDWs are
polarized in the $(x,y)$-plane. $\phi _N$ and $\phi _{-N}$ determine the
direction of the spin magnetization, and $\theta _N$ and $\theta _{-N}$ the
position of the SDWs with respect to the crystal lattice. The ratio $|\gamma
|=m_{-N}/m_N$ of the amplitudes of the two SDWs in the Ribault phase is not
precisely known. It depends on the detailed geometry of the Fermi surface
(via $t_{4b}$) and increases with decreasing pressure. However, the stability of the
Ribault phase against the formation of two helicoidal (i.e., circularly
polarized) SDWs requires $|\gamma | \stackrel{\textstyle <}{\sim }
0.5$ \cite{ND}.  

The FISDW phases are characterized by the complex order parameters
$\Delta _{\beta N,\alpha}$ defined by
\begin{equation}
\Delta _\alpha ({\bf r})=\langle {\psi}^\dagger _{-\alpha,\downarrow}
({\bf r}){\psi}_{\alpha,\uparrow}({\bf r}) \rangle
=\sum_{\beta =\pm}\Delta _{\beta N,\alpha}
e^{i\alpha{\bf r}\cdot{\bf Q}_{\beta N}}.
\label{op}
\end{equation}
The operators $\psi _{\alpha ,\sigma }^{(\dagger )} ({\bf r})$ annihilate
(create) electrons with spin $\sigma $ and momenta close to
$\alpha k_F$ ($\alpha =\pm $). The order parameters are entirely determined
by the mean value of the spin density (Eq.\ (\ref{sdsin})). 
In particular, $|\Delta _{\beta N,+}|=|\Delta _{\beta N,-}|$ 
($\beta =\pm $) for sinusoidal (i.e., linearly polarized) SDWs \cite{ND}. 

First we consider the zero-temperature condensation energy $\Delta E$ at the
mean-field level using the quantum limit approximation (QLA), also known as 
the single gap approximation. This approximation is valid when $v_FG\gg T$
and consists in retaining only the gaps at the Fermi level, neglecting those
opening above and below the Fermi level \cite{Heritier84,ND}. Introducing
\begin{equation}
\tilde \Delta_{\beta N,\alpha} = I_{\beta N}
(g_2\Delta _{\beta N,\alpha} + g_3\Delta _{-\beta N,-\alpha}),
\label{tDel}
\end{equation}
we write the condensation energy as 
\begin{equation}
\Delta E = \sum _\alpha  \Biggl \lbrace \sum _\beta 
\frac{\Delta _{\beta N,\alpha }^*\tilde \Delta _{\beta N,\alpha }}{I_{\beta
N}} +\frac{N(0)}{2}  |\tilde \Delta _{-N,\alpha }|^2  
- \frac{N(0)}{2} \sum _\beta |\tilde \Delta _{\beta N,\alpha }|^2 
\Biggl ( \frac{1}{2} +
\ln \Biggl \vert \frac{2E_0}{\tilde \Delta _{N,\alpha }} \Biggr \vert 
\Biggr ) \Biggr \rbrace .
\end{equation}
Here $E_0$ is an ultraviolet cutoff of the order of $t_a$, and 
$N(0)=1/\pi v_Fb$  the density of states per spin. The coefficients
$I_n\equiv I_n(q_yb=\pi)$ are well known in the QNM. They depend on the
transverse dispersion law $t_\perp (k_yb)$ and measure the degree of perfect
nesting of the Fermi surface \cite{Heritier84,ND}. 

Minimizing $\Delta E$ with respect to the order parameters $\Delta
_{\beta N,\alpha }$, we find that the mean-field ground state corresponds to
$\theta _N=-\theta _{-N}$ and $\phi _N=\phi _{-N}$. The latter relation 
shows that both SDWs have the same polarization axis. The condition $\theta
_N=-\theta _{-N}$ 
means that the two SDWs can be displaced in opposite directions without
changing the energy of the system. This property is related to  the
pinning that would occur for a commensurate SDW. Indeed, for a single SDW
with wave vector $(2k_F,\pi/b)$, the condition $\theta _N+\theta _{-N}=0$
becomes the usual pinning condition $\theta =0$, where $\theta $ is the
phase of the SDW \cite{Gruner}. The degeneracy of the ground state results
from rotational invariance around the $z$ axis in spin space \cite{note2} and
translational invariance in real space. The latter holds in the FISDW phases,
since the SDWs are incommensurate with respect to the crystal lattice
\cite{note3}. 

According to the mean-field analysis, and in agreement with general symmetry
considerations, we therefore expect two (gapless) Goldstone modes: 
a spin-wave mode corresponding to a uniform rotation around the $z$ axis
of the common polarization axis, and a sliding mode corresponding
to a displacement of the two SDWs in opposite directions. 
\vspace{.3cm}

\noindent
{\it Long wavelength collective modes.} - 
Collective modes can be studied by expressing the partition
function as a functional integral over bosonic fields describing spin
fluctuations \cite{Maki87}. 
The standard mean-field theory is recovered from a saddle
point approximation. Collective modes are obtained studying small (Gaussian)
fluctuations around the saddle point \cite{Fradkin}. The
static mean-field order parameters $\Delta _\alpha ({\bf r})$ (Eq.\
(\ref{op})) then become space- and time-dependent fluctuating variables
$\Delta _\alpha ({\bf r},\tau )=\Delta _\alpha ({\bf r})+\delta \Delta
_\alpha ({\bf r},\tau )$. Here $\tau $ is an
imaginary time. We use the Matsubara formalism and obtain real-time
quantities by standard analytic continuation. Taking advantage of the
decoupling of phase and amplitude modes in the long wavelength limit, we
do not consider the latter. Order parameter fluctuations $\delta
\Delta _\alpha ({\bf r},\tau )$ then correspond to fluctuations of the phase
variables $\theta _{\pm N}({\bf r},\tau )$ and $\phi _{\pm N}({\bf r},\tau
)$. Skipping technical details \cite{ND1}, we only quote the
final result for the collective modes, restricting ourselves to longitudinal
(i.e., parallel to the chains) fluctuations. 

We find two Goldstone modes with a linear dispersion law $\omega =v_Fq_x$: a
sliding mode and a spin-wave mode corresponding to
\begin{equation}
\theta _N(q_x,\omega )=-\theta _{-N}(q_x,\omega ),\,\,
\phi _N(q_x,\omega )=\phi _{-N}(q_x,\omega ),
\label{goldstone}
\end{equation}
respectively. $\theta _{\pm N}(q_x,\omega )$ and $\phi _{\pm N}(q_x,\omega )$
are the Fourier transforms of the phase variables $\theta _{\pm N}(x,t)$ and
$\phi _{\pm N}(x,t)$ (after analytic continuation to real time $t$).
Eq.\ (\ref{goldstone}) agrees with previous conclusions drawn from the
mean-field analysis. Notice that the oscillations of the two SDWs are
in-phase in the gapless spin-wave mode, and out-of-phase in the gapless
sliding mode.  

We also find a gapped sliding mode, $\omega ^2=v_F^2q_x^2 +\omega _1^2$, and
a gapped spin-wave mode, $\omega ^2=v_F^2q_x^2 +\omega _2^2$, with: 
\begin{equation}
\omega _1^2 =
\frac{12}{g_2N(0)} \frac{r}{1-r^2} 
\frac{|\tilde \Delta _{N,+}\tilde \Delta _{-N,+}|}{|I_NI_{-N}|} 
\frac{3+5\tilde \gamma ^2}{3\tilde \gamma ^2}, \,\,\,
\omega _2^2 =
\frac{12}{g_2N(0)} \frac{r}{1-r^2} 
\frac{|\tilde \Delta _{N,+}\tilde \Delta _{-N,+}|}{|I_NI_{-N}|} 
\frac{1-\tilde \gamma ^2}{\tilde \gamma ^2},
\label{gapped}
\end{equation}
where $r=g_3/g_2$ and $\tilde \gamma =\tilde \Delta _{-N,+}/\tilde \Delta
_{N,+}$. $\tilde \gamma $ is
related to $\gamma =\Delta _{-N,+}/\Delta _{N,+}$ ($|\gamma |=m_{-N}/m_N$) 
by $\gamma =(\tilde \gamma I_N-rI_{-N})/(I_{-N}-r\tilde 
\gamma I_N)$. In the Ribault phase, $|\tilde \gamma |\simeq |\gamma |$
\cite{ND}. With no loss of
generality, we choose $\Delta _{\beta N,\alpha }$ and 
$\tilde \Delta _{\beta N,\alpha }$ to be real. The gapped sliding mode
corresponds to a displacement of the two SDWs in the same direction (i.e.,
${\rm sgn}[\theta _N(x,t)]={\rm sgn}[\theta _{-N}(x,t)]$), while the gapped
spin-wave mode corresponds to opposite rotations of the SDWs  (i.e.,
${\rm sgn}[\phi _N(x,t)]=-{\rm sgn}[\phi _{-N}(x,t)]$). In particular, when
$|\tilde \gamma |=|\gamma |=1$ (a situation reached when $|I_N|=|I_{-N}|$), 
gapped modes correspond to $\theta _N=\theta _{-N}$ and
$\phi _N=-r\phi _{-N}$. Using the physical parameters of the Bechgaard
salts, we find that $\omega _1$ and $\omega _2$ are larger than
the mean-field order parameters $|\tilde \Delta _{\pm N,\pm}|$, so that the
gapped modes appear above the quasi-particle excitation gap (generally within
the first Landau subband above the Fermi level). Therefore, we expect these
modes to be strongly damped due to the coupling with quasi-particle
excitations.   

There are some similarities between collective modes of the SDWs and
phase modes occuring in two-band or bilayer superconductors
\cite{Legett66,Wu95}. $g_3/g_2$ plays the same role as the ratio between the
intraband (or intralayer) and interband (or interlayer) coupling constants. 
To some extent, there are also analogies with plasmon modes occuring in
conducting bilayer systems \cite{Dassarma81}. While the corresponding phase
modes in superconducting systems have not yet been observed, plasmon
modes in semiconductor double-well structures have been observed recently
via inelastic light scattering experiment \cite{Kainth98}.
\vspace{.3cm}

\noindent
{\it Spectral functions.} -
Now we consider the spectral functions of the collective modes of the SDWs:
the spin-spin correlation
function $\chi _{yy}^{\rm ret}({\bf q},{\bf q}',\omega)$ and the optical
conductivity $\sigma ({\bf q},\omega )=(i/\omega )\Pi^{\rm ret}({\bf
q},\omega )$, where $\Pi^{\rm ret}$ is the current-current correlation
function. $\Pi^{\rm ret}$ and $\chi _{yy}^{\rm ret}$ are the retarded
parts of the imaginary time correlation functions
\begin{equation}
\Pi({\bf r},\tau ;{\bf r}',\tau ')= \langle j_{\rm DW}({\bf
r},\tau) j_{\rm DW}({\bf r}',\tau ')\rangle , \,\,\,
\chi _{yy}({\bf r},\tau ;{\bf r}',\tau ')= \langle  S_y({\bf
r},\tau) S_y({\bf r}',\tau ')\rangle .
\label{cf}
\end{equation}
$j_{\rm DW}$ is the current along the chains carried by SDW fluctuations. 
For real $\Delta _{\beta N,\alpha }$, the mean-field magnetization (Eq.\
(\ref{sdsin})) is along the $x$ axis so that $S_y$ corresponds to transverse
spin fluctuations. To lowest order in phase fluctuations \cite{ND1}
\begin{eqnarray}
j_{\rm DW}({\bf r},\tau)&=&-\frac{ie}{\pi b}
\frac{\partial }{\partial \tau }
\frac{\theta _N({\bf r},\tau)-r\gamma \theta _{-N}({\bf r},\tau)} 
{1+r\gamma } , \nonumber \\ 
S_y({\bf r},\tau)&=& 2\sum _\beta \Delta _{\beta N,+}\cos ({\bf r}\cdot {\bf
Q}_{\beta N}) \phi _{\beta N}({\bf r},\tau). 
\label{Sy}
\end{eqnarray}
$j_{\rm DW}$ is a function of $\theta _{\pm N}$, while $S_y$ is a function
of $\phi _{\pm N}$. Thus, $\Pi$ is determined by the sliding modes, while
$\chi _{yy}$ is determined by the spin-wave modes. 

In the limit ${\bf q}=0$, the dissipative part of the conductivity is given 
by \cite{note}
\begin{equation}
{\rm Re}[\sigma (\omega )] = \frac{\omega _p^2}{4} \Biggl ( \delta
(\omega ) \frac{3(1-\tilde \gamma ^2)}{3+5\tilde \gamma ^2}
+ \delta (\omega \pm \omega _1)
\frac{4\tilde \gamma ^2}{3+5\tilde \gamma ^2} \Biggr ). 
\label{opt}
\end{equation}
We have introduced the plasma frequency $\omega _p=\sqrt{8e^2v_F/b}$. 
Eq.\ (\ref{opt}) satisfies the conductivity sum rule $\int _{-\infty}
^{\infty }d\omega \, {\rm Re}[\sigma (\omega )]=\omega
_p^2/4$. Quasi-particle excitations above the mean-field gap do not
contribute to the optical conductivity, a result well known in SDW systems
\cite{Gruner}. Because both modes contribute to the 
conductivity, the low-energy (Goldstone) mode carries only a fraction of the
total spectral weight. We obtain Dirac
peaks at $\pm \omega _1$ because we have neglected the coupling of the gapped
mode with quasi-particle excitations. Also, in a real system
(with impurities), the Goldstone mode would broaden and appear at a finite
frequency (below the quasi-particle excitation gap) due to pinning by
impurities. In the clean limit, which is appropriate in (TMTSF)$_2$X salts,
the presence of impurities does not restore any significant spectral weight to
quasi-particle excitations above the mean-field gap \cite{Gruner}. 
Therefore, the fraction of spectral weight carried by the two
modes is correctly given by Eq.\ (\ref{opt}). By measuring the optical
conductivity $\sigma (\omega )$, we can therefore obtain the ratio $|\tilde
\gamma |\simeq |\gamma |$ of the amplitudes of the two SDWs. We have shown
in Ref.\ \cite{ND} that $|\tilde \gamma |$ can vary between $\sim 0$
and $\sim 0.5$ in the Ribault phase.   
Thus, the Goldstone mode can carry between $\sim 100\%
$ and $\sim 50\% $ of the total spectral weight. When this fraction is reduced
below $\sim 50\% $ (i.e., when $|\tilde \gamma | 
\stackrel{\textstyle >}{\sim } 0.5$), the
Ribault phase becomes unstable against the formation of a helicoidal
phase \cite{ND}. The low-energy spectral weight is therefore also a measure
of the stability of the sinusoidal Ribault phase against the formation of a
helicoidal phase. For a helicoidal structure, we cannot distinguish between
a uniform spin rotation and a global translation, so that there is only one
type of modes. Thus, in the helicoidal phase, we find a gapless Goldstone
mode and a gapped mode. The Goldstone mode carries no current, and all the
spectral weight is pushed up above the quasi-particle excitation gap
\cite{ND1}.  

Due to the presence of SDWs, the spin-spin correlation function $\chi
_{yy}^{\rm ret}$ is not diagonal in momentum space, but has components $\chi
_{yy}^{\rm ret}({\bf q}+\alpha {\bf Q}_{\beta N},{\bf q}+\alpha '{\bf
Q}_{\beta ' N},\omega )$ where $\alpha,\alpha',\beta,\beta'=\pm$ (see Eqs.\
(\ref{cf}--\ref{Sy})). ${\bf q}$ corresponds to the momentum of the
spin-wave mode and tends to zero for long wavelength fluctuations. We
therefore consider the spectral function ${\rm Im\,Tr}_{\bf q}\chi_{yy}
^{\rm ret}$, where ${\rm Tr}_{\bf q}$ is a partial trace corresponding to a
given spin-wave mode momentum $\bf q$. In the limit $q_y=0,q_x\to 0$, we
obtain \cite{note}
\begin{eqnarray}
{\rm Im}{\rm Tr}_{\bf q}\chi _{yy}^{\rm ret}&=& \frac{2\pi }{g_2^2N(0)} 
\frac{|\tilde \Delta _{N,+}\tilde \Delta _{-N,+}|} {|I_NI_{-N}|}
\Biggl \lbrace \frac{\delta (\omega -v_Fq_x)}{v_Fq_x}
\Biggl \lbrack \frac{-4r}{(1-r^2)^2}+\frac{(1+r^2)}{(1-r^2)^2}
 \frac{\tilde \gamma ^2+\zeta^2}{|\tilde \gamma \zeta|} \Biggr \rbrack  
\nonumber \\ && +
\frac{\delta (\omega -\omega _2)}{\omega _2} \frac{3(1+r^2)}{2(1-r^2)^2}
\frac{1-\tilde \gamma ^2}{|\tilde \gamma \zeta|} \Biggr \rbrace , 
\label{Cyy}
\end{eqnarray}
for $\omega ,q_x>0$. $\zeta=I_{-N}/I_N$. Both spin-wave modes
contribute to the spectral function. The spectral weight carried by the
Goldstone mode diverges as $1/q_x$ as expected for a quantum antiferromagnet
\cite{Fradkin}. Eqs.\ (\ref{opt}) and (\ref{Cyy}) predict that all the spectral
weight is carried by the in-phase modes, i.e., the gapped sliding mode and
the gapless spin-wave mode, whenever both SDWs have the same amplitude
($|\tilde \gamma |=|\gamma |=|\zeta |=1$).
\vspace{.3cm}

\noindent
{\it Conclusion.} - We have shown that the Ribault anomaly in the FISDW phases
of the Bechgaard salts \cite{Ribault85} is characterized not only by a sign
reversal of the QHE, but also by a 
rich structure of collective sliding and spin-wave modes. The presence of
two SDWs in this phase gives rise to in-phase and out-of-phase
collective oscillations in the long wavelength limit. The out-of-phase sliding
and in-phase spin-wave modes are gapless (Goldstone modes). The in-phase
sliding and out-of-phase spin-wave modes are gapped and are expected to
appear above the quasi-particle excitation gap in the Ribault phase. 
Charge and spin response functions are strongly affected by the presence of
these modes. In particular, we have shown that the low-energy (Goldstone)
sliding mode carries only a fraction of the total spectral weight in the
optical conductivity. By measuring the latter, we can obtain the ratio of 
the amplitudes of the two
SDWs that coexist in the Ribault phase. The low-energy spectral weight is
also a measure of the stability of the Ribault phase against the formation
of helicoidal SDWs. Helicoidal SDWs have been predicted in Ref.\
\cite{ND} and are expected to appear at low pressure (since $|\tilde \gamma
|$ increases with decreasing pressure). We have
pointed out some analogies with collective modes in two-band
superconductors or bilayer systems \cite{Legett66,Wu95,Dassarma81}. The
common feature to all these collective modes is that they appear in
a two-component system. To our knowledge, it is the first time that such
modes are predicted in SDW systems. 

We thank S. Das Sarma for useful discussions and for drawing our attention
to collective modes in double-layer semiconductor structures. This work was
partially supported by the NSF under Grant DMR--9417451 and by the Packard
Foundation.


\begin{thebibliography}{99}
\bibitem{coex} Because of umklapp scattering, all FISDW phases contain two
SDWs. However, the two SDWs have comparable amplitudes only in the Ribault
phase. In the positive phases, one of the SDWs has a vanishing
amplitude and can therefore be neglected \cite{ND}.
\bibitem{note1} The reason is that the harmonics $t_{mb}$ ($m=1,2...$)
enters the mean-field theory only through the coefficients $I_n(q_yb)$
introduced in Eq.\ (\ref{tDel}). These coefficients do not depend on
the odd harmonics $t_{(2m+1)b}$ for $q_yb=\pi$ \cite{Zanchi96}. In the
Ribault  phase, only the case $q_yb=\pi$ arises.
\bibitem{note} Note that the long-range Coulomb interaction (not considered
here) affects neither the transverse sliding modes sampled by optical 
spectroscopy nor the spin-wave modes.
\end{thebibliography}
\end{document}